\documentclass[12pt,a4paper]{article}
\usepackage{latexsym}
\def\be{\begin{equation}}
\def\ee{\end{equation}}
\def\bea{\begin{eqnarray}}
\def\eea{\end{eqnarray}}

\def\pd{\partial}

\def\eps{\epsilon}
\def\dl{\delta}
\def\tldGm{\tilde \Gamma}
\def\tldnb{\tilde \nabla}

\def\vmu{\vec\mu}
\def\vmup{\vec\mu'}
\def\vmpp{\vec\mu''}
\def\vnu{\vec\nu}
\def\vnup{\vec\nu'}

\def\vrho{\vec\rho}

\def\vsig{\vec\sigma}

\def\kp{\kappa}
\def\vkap{\vec\kappa}
\def\vkp{\vec\kappa'}
\def\vkpp{\vec\kappa''}
\def\vtau{\vec\tau}
\def\vtap{\vec\tau'}
\def\veta{\vec\eta}

\def\lm{\lambda}
\def\sig{\sigma}
\begin{document}
\begin{flushright}
DTP/00/93
\end{flushright}
\vspace{0.5cm}
\begin{center}
\begin{Large}
{\bf Covariant Formulation of Field Theories
associated with $p$-Branes}
\end{Large}
\\
\vspace{1.0cm}
{\bf David B. Fairlie$\footnote{e-mail: David.Fairlie@durham.ac.uk}$ 
and Tatsuya Ueno$\footnote{e-mail: Tatsuya.Ueno@durham.ac.uk}$}\\
\vspace{1.0cm}
Department of Mathematical Sciences,\\
         Science Laboratories,\\
         University of Durham,\\
         Durham, DH1 3LE, England \\
\vspace{0.7cm}
9th November 2000
\end{center}
\vspace{0.8cm}
\begin{abstract} 
We discuss the covariant formulation of local field theories 
described by the Companion Lagrangian associated with $p$-branes. 
The covariantisation is shown to be useful for clarifying the 
geometrical meaning of the field equations and also their relation 
to the Hamilton-Jacobi formulation of the standard Dirac-Born-Infeld 
theory.
\end{abstract}

\section{Introduction}%
A significant class of equations of motion occurring in Physics have 
the property of general covariance; i.e the property that the solutions 
of these equations remain solutions under a large set of 
transformations. 
The best known examples are the equations of General Relativity, 
Yang-Mills and the Maxwell equations in terms of the gauge fields. 
The fact, realised by the cognoscenti, that the origin of covariance 
in those examples could be unified by the construction of a covariant 
derivative became generally known in the mid 1970's with the adoption 
of fibre bundle language in the discussion of connections. 
In this article we discuss a further example of the equations of 
motion arising from what we have called the Companion Lagrangian, 
which may be considered as a continuation of the Dirac-Born-Infeld 
(DBI) equations describing $D$-branes to the situation where the 
target space is of smaller dimension than the base space. 
The genesis of these equations lies in the idea of replicating for 
strings and branes the situation in ordinary quantum mechanics in 
which the classical point particle Lagrangian is replaced by the 
quantum Klein Gordon Lagrangian \cite{dbf1,dbf2,dbf3,hos1,hos2}.
\par

Let $\phi^i$ be $n$ fields each dependent upon co-ordinates $x^\mu$ 
($\mu=1,2,\ldots,d > n$) of the base space. 
Then in its simplest form the Companion Lagrangian ${\cal L}$ is
\be
{\cal L}=\sqrt{\det\left|\frac{\pd \phi^i}{\pd x_\mu}
\frac{\pd \phi^j}{\pd x_\mu}\right|} \ .
\label{companion}
\ee
The simplest example of an equation arising from this for $n=1,\ d=2$ is 
the Bateman equation,
\be
\left(\frac{\pd \phi}{\pd x_1}\right)^2\frac{\pd^2 \phi}{\pd x_2^2}+
\left(\frac{\pd \phi}{\pd x_2}\right)^2\frac{\pd^2 \phi}{\pd x_1^2}-2
\left(\frac{\pd \phi}{\pd x_1}\right)\left(\frac{\pd \phi}{\pd 
x_2}\right)
\frac{\pd^2 \phi}{\pd x_1\pd x_2}\,=\,0 \ ,
\label{bateman}
\ee
\cite{batman,gov} and for $n=1$, arbitrary $d$ the Companion equation 
is a sum of $d \choose 2$ such Bateman expressions set to zero. 
It is known that the solution of (\ref{bateman}) is given implicitly 
by solving the equation,
\be
x_1 F(\phi(x)) + x_2 G(\phi(x)) = c \ ,
\label{bateman-sol}
\ee
where $F, G$ are arbitrary functions and $c$ is a constant. 
{}From the form of the solution, it is obvious that the Bateman 
equation is invariant under any change of $\phi \rightarrow \phi'(\phi)$. 
Another example is for two fields $\phi,\ \psi$ in three dimensions, 
in which the Companion equation can be recast in the form,
\be
\det\left|\begin{array}{ccccc}
0 & 0 & \phi_1 & \phi_2 & \phi_3\\
0 & 0 & \psi_1 & \psi_2 & \psi_3\\
\phi_1 & \psi_1 & \phi_{11} & \phi_{12} & \phi_{13}\\
\phi_2 & \psi_2 & \phi_{12} & \phi_{22} & \phi_{23}\\
\phi_3 & \psi_3 & \phi_{13} & \phi_{23} & \phi_{33}
\end{array}\right|=0 \ , \label{gb3}
\ee
with a similar equation with second derivatives of $\psi$, where 
subscripts denote derivatives, e.g., $\phi_{\mu} = \pd \phi/\pd x^{\mu}$.
These equations have been studied in \cite{gov} as the one of the Universal 
Field Equations and shown to be covariant under any arbitrary redefinition 
of the fields.
This remarkable property also holds for the general $(n,d)$ case, 
that is, {\it the Companion Equation is invariant under any change of 
fields $\phi^i \rightarrow \phi'^i(\phi^j)$}, which will be transparent in 
the following covariant formulation.
This symmetry  corresponds to the general reparametrization invariance in 
the DBI formulation of $p=n-1$ branes, where the theory with $d$ fields 
$X^{\mu}(\tau^i)$ is invariant under reparametrization of the $n$ 
worldvolume co-ordinates $\tau^i$. 
\par

In the next section, we reconsider the Companion theory in a 
manifestly covariant way and clarify the geometrical meaning of the 
Companion equations.
In the section 3, we study the relation between the Hamilton-Jacobi 
formulation of the DBI theory and the Companion theory, showing that 
the latter possesses a class of solutions of the former, which are 
characterized by a divergence-free condition of a degenerate metric
defined in the latter theory.
The relation between the DBI and Companion theories is demonstrated
in the section 4 explicitly in the particle ($n=1$) case in an arbitrary 
number of dimensions.
\par

\section{Covariant Formulation}%
\subsection{Notation}%
For simplicity, we work in $d$-dimensional Euclidean space with 
the flat metric $\dl_{\mu\nu}$ and the totally antisymmetric tensor
$\eps_{\nu_1\nu_2\ldots\nu_d}$ with $\eps_{12\ldots d}=+1$.
Indices with an arrow above them represent the set of several 
indices. 
$\vmu,\vnu,\vrho,\vsig$ each have $n$ components, e.g., 
$\vmu=\{\mu_1,\mu_2,\ldots,\mu_n\}$. 
$\vtau, \vkap, \veta$ each have $(d-n)$ components, e.g., 
$\vkap = \{\kp_1, \kp_2,\ldots,\kp_{d-n}\}$.
If the (double) prime $'$ ($''$) is used for the indices, $\vmu,\vnu$ 
or $\vtau,\vkap$, then their components start from the second (third) 
entry of un-primed ones, e.g., $\vmup=\{\mu_2,\ldots,\mu_n\}$,
$\vkp=\{\kp_2,\ldots,\kp_{d-n}\}$, $\vmpp=\{\mu_3,\ldots,\mu_n\}$
and $\vkpp=\{\kp_3,\ldots,\kp_{d-n}\}$.

\subsection{Jacobians}%
The Companion equations may be expressed more succinctly in terms of 
the Jacobians, which are defined as
\bea
J_{\vkap}=J_{\kp_1\kp_2\ldots\kp_{d-n}}
&&=\eps_{\kappa_1\kappa_2\ldots\kappa_{d-n}\nu_1\nu_2\ldots\nu_n}
\phi^1_{\nu_1}\phi^2_{\nu_2}\ldots\phi^n_{\nu_n} 
\nonumber \\
&&= {1 \over n!}\, \eps_{\vkap\vnu}\, \eps_{i_1\ldots i_n}
\phi^{i_1}_{\nu_1}\ldots\phi^{i_n}_{\nu_n}
\equiv {1 \over n!} \, \eps_{\vkap\vnu} \, {\tilde J}_{\vnu} \ .
\label{j}
\eea
The derivatives of the Jacobians are
\be
{\pd J_{\vkap} \over \pd \phi^i_{\mu}} 
= {1 \over (n-1)!} \, \eps_{\vkap\mu\vnup} \, \eps_{ii_2\ldots i_n}
\phi^{i_2}_{\nu_2}\ldots\phi^{i_n}_{\nu_n}
\equiv {1 \over (n-1)!} \, \eps_{\vkap\mu\vnup}\, {\tilde J}_{i,\vnup} 
\ .  \label{de1j}
\ee
Using the Jacobians, the Companion Lagrangian (\ref{companion}) is 
written as
\be
{\cal L}\,=\, \sqrt{\det
\left|\frac{\pd\phi^i}{\pd x_\mu}\frac{\pd\phi^j}{\pd x_\mu}\right|}
\,=\, \sqrt{{1 \over (d-n)!} J_{\vkap} J_{\vkap}} 
\,=\, \sqrt{{1 \over n!} {\tilde J}_{\vmu} {\tilde J}_{\vmu}} \ ,
\label{com2}
\ee
from which the equation of motion is derived,
\be
\frac{\pd^2 {\cal L}}{\pd \phi^i_\mu \pd \phi^j_\nu} \phi^j_{\mu \nu} 
= {1 \over (d-n)!^2} {\cal L}^{-3} 
(J_{\vtau}J_{\vtau}{\pd J_{\vkap} \over \pd \phi^i_{\mu}}
{\pd J_{\vkap} \over \pd \phi^j_{\nu}} 
- J_{\vkap} {\pd J_{\vkap} \over \pd \phi^i_{\mu}}
J_{\vtau} {\pd J_{\vtau} \over \pd \phi^j_{\nu}})\, \phi^j_{\mu \nu} 
= 0 \ . \label{emsq}
\ee
As shown in \cite{lb}, using the identity of epsilon tensors, 
\bea
\eps_{\mu\nu_2\nu_3\ldots\nu_d}\eps_{\rho_1\rho_2\ldots\rho_d}
=\eps_{\rho_1\nu_2\nu_3\ldots\nu_d}\eps_{\mu\rho_2\ldots\rho_d}
+\eps_{\rho_2\nu_2\nu_3\ldots\nu_d}\eps_{\rho_1\mu\rho_3\ldots\rho_d}
+\ldots\nonumber\\ 
\ldots 
+\eps_{\rho_d\nu_2\nu_3\ldots\nu_d}\eps_{\rho_1\rho_2\ldots\rho_{d-1}\mu} 
\ ,
\label{epid}
\eea
where the index $\mu$ is swapped with each index in the second 
epsilon, 
we can rewrite the equation as
\be
\frac{\pd^2 {\cal L}}{\pd \phi^i_\mu \pd \phi^j_\nu} \phi^j_{\mu \nu}
= {1 \over {(r+1)!(r-1)!}} {\cal L}^{-3}({\pd J_{\vkap} \over \pd\phi^i_\tau}
{\pd J_{\vkap} \over \pd \phi^j_\tau})J_{\mu\vtap}J_{\nu\vtap}\phi^j_{\mu\nu}
= 0 \ ,
\ee
where $r=d-n$. 
Under the assumption $\det|\pd J_{\vkap}/\pd \phi^i_\tau \pd 
J_{\vkap}/\phi^j_\tau|\not= 0$, we obtain the Companion equation,
\be
J_{\mu\vkp} J_{\nu\vkp} \phi^i_{\mu \nu}=0 \ .
\label{jsq}
\ee
This equation may be interpreted as a sum of Universal Field Equations
\cite{gov}.
It will be shown below that the left-hand-side (LHS) of this equation 
appears as a covariant derivative acting on the field $\phi^i_{\mu}$.
\par

For later use, we note two useful identities for the Jacobians,
\bea
&&{\pd J_{\vkap} \over \pd \phi^j_{\mu}} \phi^i_{\mu} = \dl^i_j \, 
J_{\vkap}
\ , \label{jid-1}\\
&&{\pd J_{\vkap} \over \pd \phi^j_{\nu}} \phi^j_{\mu} = 
(r+1)\dl_{\mu[\nu}J_{\vkap]} = 
(\dl_{\mu\nu}J_{\vkap} - \dl_{\mu\kp_1}J_{\nu\kp_2\ldots\kp_r} - 
\ldots
- \dl_{\mu\kp_r}J_{\kp_1\ldots\kp_{r-1}\nu}) \, . 
\label{jid-2}
\eea
{}From the identities, we obtain 
\be
{\pd {\cal L} \over \pd \phi^j_{\mu}}\phi^i_{\mu} =
\dl^i_{\, j}{\cal L} \ , \qquad \ \ 
{\pd {\cal L} \over \pd \phi^j_{\nu}}\phi^j_{\mu} = {\cal L} \, 
(\dl_{\mu\nu} - {1 \over (r-1)!} {\cal L}^{-2} 
J_{\mu\vkp}J_{\nu\vkp}) \ .
\ee

\subsection{Induced Metric} \label{sec-im}%
The field $\phi^i(x)$ is the mapping from the $d$-dimensional base 
space with the flat metric $\dl_{\mu\nu}$ to the $n$-dimensional space 
labelled by $\{\phi^i\}$.
The induced metric $G^{ij}$ on the $\phi$-space is then defined as 
the pullback of $\dl^{\mu\nu}$,
\be
G^{ij} = \phi^i_{\mu}\phi^j_{\mu} \ .
\label{im}
\ee
The metric 
$G^{ij} \rightarrow (\pd \phi'^i/\pd \phi^k)(\pd \phi'^j/\pd \phi^l)G^{kl}$ 
transforms under the reparametrization of $\phi^i$, as anticipated.
The inverse of $G^{ij}$ can be obtained explicitly by using the identities 
(\ref{jid-1}) and (\ref{jid-2}),
\be
G_{ij} = {1 \over (r+1)!} {\cal L}^{-2}
{\pd J_{\vkap} \over \pd \phi^i_{\mu}}{\pd J_{\vkap} \over \pd 
\phi^j_{\mu}}
= {1 \over (n-1)!} {\cal L}^{-2} {\tilde J}_{i,\vmup} {\tilde J}_{j,\vmup} 
\ . \label{iim}
\ee
Note that the square of the Lagrangian can be written as 
${\cal L}^2 = \det|G^{ij}|$.
Having obtained $G_{ij}$, we go back to the base space again with the 
induced metric $g_{\mu\nu}$, the pullback of $G_{ij}$,
\bea
g_{\mu\nu} = \phi^i_{\mu}\phi^j_{\nu} G_{ij} &=& {1 \over (n-1)!}
{\cal L}^{-2}{\tilde J}_{\mu\vmup}{\tilde J}_{\nu\vmup}
\nonumber \\
&=& \dl_{\mu\nu} - {1 \over (r-1)!} {\cal 
L}^{-2}J_{\mu\vkp}J_{\nu\vkp}
\ ,
\label{deg}
\eea
where the identity (\ref{jid-2}) has been used in the last line.
The metric $g_{\mu\nu}$ is manifestly invariant under the reparametrization 
of $\phi^i$, although it is degenerate in our $d > n$ case. 
Note that $g_{\mu\nu}$ is the flat metric $\dl_{\mu\nu}$ for $d=n$, 
while it cannot be defined for $d < n$.
Let us write down the metrics explicitly in the particle ($n=1$) and 
string ($n=2$) cases,
\bea
&&n=1, \quad \ g_{\mu\nu} = 
{\phi_{\mu}\phi_{\nu} \over \phi_{\lm}\phi_{\lm}} \ , \\
&&n=2, \quad \ g_{\mu\nu} = 
{{\tilde J}_{\mu\rho}{\tilde J}_{\nu\rho} \over 
{\det |\phi^i_{\lm}\phi^j_{\lm}|}} 
= - {\cal L}^{-2} \left|\begin{array}{ccc}
 0 & \phi^1_\nu & \phi^2_\nu\\
\phi^1_\mu & \phi^1_\lm \phi^1_\lm & \phi^1_\lm \phi^2_\lm \\
\phi^2_\mu & \phi^2_\lm \phi^1_\lm & \phi^2_\lm \phi^2_\lm 
\end{array}\right| \ . 
\eea
In these expressions, it is easy to see that $\phi^i_{\mu}$ are 
eigenvectors of $g_{\mu\nu}$ with eigenvalue $+1$.
This important property of $g_{\mu\nu}$ holds for general $(n,d)$,
\be
g_{\mu\nu} \phi^i_{\nu} = \phi^j_{\mu} G_{jk} G^{ki} = \phi^i_{\mu} \ ,
\label{eigen}
\ee
which leads us to define the projection operator
$P_{\mu\nu}= (\dl_{\mu\nu} - g_{\mu\nu})$ acting on the 
$d$-dimensional vector space ${\cal V}$ in the base space. 
Then ${\cal V}$ is decomposed as the sum of two subspaces ${\cal V}_n$ and 
${\cal V}_r$.
The latter is the $(d-n)$-dimensional subspace orthogonal to the $n$ 
vectors $\phi^i_{\mu}$, while the former is spanned by a linear combination 
of $\phi^i_{\mu}$.
For an arbitrary $V_{\mu} \in {\cal V}$, we have
\be
V_{\mu} = g_{\mu\nu}V_{\nu} + P_{\mu\nu}V_{\nu} = 
V_i \, \phi^i_{\mu} + P_{\mu\nu}V_{\nu} \ .
\label{vector}
\ee
We introduce the dual vectors $Y_{j\nu} = {\cal L}^{-1}
(\pd {\cal L}/\pd \phi^j_{\nu}) =G_{jk}\phi^k_{\nu}$ of 
$\phi^i_{\mu}$, which satisfy
\be
\phi^i_{\mu} \, Y_{j\mu} = \dl^i_j \ , \qquad
\phi^i_{\mu} \, Y_{i\nu} = Y_{i\mu} \phi^i_{\nu} = g_{\mu\nu} \ ,
\label{dualY}
\ee
then the component $V_i$ of the vector $V_{\mu}$ in (\ref{vector}) is 
expressed as $V_i = V_{\mu} Y_{i\mu}$.
We will see in the next subsection that the connection for the 
reparametrization $\phi^i \rightarrow \phi'^i$ is constructed in terms of 
the fields $\phi^i_{\mu}$ and $Y_{j\nu}$.

\subsection{Induced Connection}%
To construct the covariant formulation under the transformation 
$\phi^i \rightarrow \phi'^i(\phi^j)$, we first consider the geometry 
of the $n$-dimensional space labelled by co-ordinates $\{\phi^i\}$, with 
the metric $G^{(0)}_{ij}(\phi)$.
The standard way to build covariant equations is to use the covariant 
derivative $\nabla_j$, with a connection $\Gamma^i_{jk}$,
\be
\nabla_j V^i = {\pd V^i \over \pd \phi^j} \, + \, \Gamma^i_{jk}V^k \ .
\ee
The assumption of the covariant constancy of the metric, 
$\nabla_jG^{(0)}_{ik}=0$, and the torsion-free condition give
the form of a Christoffel symbol to the connection $\Gamma^i_{jk}$.
The pullback of $\Gamma^i_{jk}$ onto the base space is obtained by 
acting with the factors $\phi^j_{\nu}\phi^k_{\mu}\phi^l_{\lm}$ on it and 
using the relation $\phi^j_{\nu}\pd_j = \pd_{\nu}$,
\be
(\phi^m_{\nu\mu} + \phi^j_{\nu}\phi^k_{\mu}\Gamma^m_{jk})\,
\phi^l_{\lm} G^{(0)}_{lm} 
= {1 \over 2}(\pd_{\nu}g^{(0)}_{\lm\mu} + \pd_{\mu}g^{(0)}_{\lm\nu} 
- \pd_{\lm}g^{(0)}_{\nu\mu})
\equiv \tldGm_{\lm\nu\mu}(g^{(0)}) \ ,
\label{con}
\ee
where $g^{(0)}_{\lm\mu} = \phi^l_{\lm}\phi^k_{\mu}G^{(0)}_{lk}$.
The substitution of the induced metric $G^{ij}=\phi^i_{\mu}\phi^j_{\mu}$ 
into (\ref{con}) leads us to the connection ${K_{\nu}}^i_{\ k}$,
\be
{K_{\nu}}^i_{\ k} = \phi^j_{\nu}\Gamma^i_{jk}|_{G^{(0)}=G} 
= - Y_{k\mu}\tldnb_{\nu}\phi^i_{\mu} \ ,
\label{K}
\ee
where $\tldnb_{\nu}\phi^i_{\mu} = \phi^i_{\nu\mu} 
- \tldGm_{\lm\nu\mu}(g) \, \phi^i_{\lm}$, with $g_{\mu\nu}$, the 
degenerate metric in (\ref{deg}).\\
Then the induced derivative $\nabla_{\nu}V^i = 
\pd_{\nu}V^i +{K_{\nu}}^i_{\ k}V^k$ is manifestly covariant under the 
reparametrization of $\phi^i$.
We, however, note that the substitution $G^{(0)}_{ij}=G_{ij}$ in 
(\ref{K}) cannot be justified since the induced metric $G^{ij}$ is not 
a function of $\phi^i$ but $\phi^i_{\mu}$, where the relation 
$\phi^j_{\nu}\pd_j = \pd_{\nu}$ used above is incorrect when it acts 
on $G^{ij}$.
Strictly speaking, the connection ${K_{\nu}}^i_{\ k}$ is {\it not} 
derived from $\Gamma^i_{jk}$ in the $\phi^i$-space but is {\it defined} 
in the analogy with $\phi^j_{\nu}\Gamma^i_{jk}(G^{(0)})$.
Another remark is that $\tldGm_{\lm\nu\mu}(g)$ in (\ref{K}) looks like a 
connection in the base space.
In fact, it can be shown that the derivative $\tldnb_{\nu}\phi^i_{\mu}$ 
behaves as a tensor under the reparametrization 
$x^{\mu}\rightarrow x'^{\mu}$ preserving the form of the flat metric 
$\dl_{\mu\nu}$; 
$\pd_{\lm}x'^{\mu}\pd_{\rho}x'^{\nu}\,\dl_{\lm\rho} = \dl_{\mu\nu}$.
\par

The covariant derivative $\nabla_{\nu}$ acting on $\phi^i_{\mu}$ 
becomes
\be
\nabla_{\nu}\phi^i_{\mu} - \tldGm_{\lm\nu\mu}\phi^i_{\lm} 
= \pd_{\nu}\phi^i_{\mu} + {K_{\nu}}^i_{\ k}\phi^k_{\mu} 
- \tldGm_{\lm\nu\mu}\phi^i_{\lm}
= P_{\mu\lm}\tldnb_{\nu}\phi^i_{\lm} \ ,
\label{covphi}
\ee
or equivalently, using 
$P_{\mu\lm}\phi^i_{\lm\nu} = \pd_\nu g_{\mu\lm}\phi^i_\lm$, we have
\be
\nabla_\nu \phi^i_\mu = 
\Gamma_{\lm\nu\mu} \phi^i_{\lm} \ , \qquad
\Gamma_{\lm\nu\mu} = \pd_\nu g_{\mu\lm} + 
g_{\mu\rho}\tldGm_{\lm\nu\rho} \ .
\label{covphi2}
\ee
The right-hand-side (RHS) of (\ref{covphi}) vanishes when it acts upon
$\phi^j_{\mu}$, which gives
\be
\nabla_{\nu}G^{ij} = 2 \phi^{(i}_{\mu}\nabla_{\nu}\phi^{j)}_{\mu}
= 2 \phi^{(i}_{\mu}\tldGm_{\lm\nu\mu}\phi^{j)}_{\lm}
= \phi^i_{\lm}\phi^j_{\mu} \pd_{\nu}g_{\lm\mu} \ ,
\label{covg}
\ee
in which the RHS is actually zero due to the identity,
\be
\pd_{\nu}g_{\lm\mu} = 
(\dl_{\lm\sig}-g_{\lm\sig})\, Y_{k\mu}\phi^k_{\sig\nu}
+(\dl_{\mu\sig}-g_{\mu\sig})\, Y_{k\lm}\phi^k_{\sig\nu} \ . 
\label{gid-1}
\ee
Hence we obtain the metricity condition $\nabla_{\nu}G^{ij}=0$ (and 
$\nabla_{\nu}G_{ij}=0$) in the base space.
The covariant derivative of $g_{\mu\nu}$ is also obtained from 
(\ref{covphi2}) and the identities $g_{\mu\rho}g_{\rho\nu}=g_{\mu\nu}$ and 
$\pd_\nu g_{\rho\lm} - \tldGm_{\rho\nu\lm} - \tldGm_{\lm\nu\rho} = 0$,
\be
\nabla_\lm g_{\mu\nu} = \Gamma_{\rho\lm\mu} g_{\rho\nu}
+ \Gamma_{\rho\lm\nu} g_{\rho\mu} = \pd_\lm g_{\mu\nu} \ ,
\ee
as anticipated since $g_{\mu\nu}$ is a scalar under the field 
redefinition of $\phi^i$.
\par

Finally, let us take the contraction $\mu=\nu$ in (\ref{covphi2}), 
then we have 
\be
\nabla_\mu\phi^i_\mu = \Gamma_{\lm\mu\mu}\, \phi^i_{\lm} =
\pd_\mu g_{\mu\lm} \, \phi^i_\lm + g_{\mu\rho}\tldGm_{\lm\mu\rho}\, 
\phi^i_\lm \ . \label{cont}
\ee
The second term of the RHS vanishes due to the identity (\ref{gid-1}),
which gives, with the formula of $g_{\mu\nu}$ in (\ref{deg}),
\be
\nabla_{\mu}\phi^i_{\mu} = \pd_{\mu}g_{\mu\lm} \, \phi^i_{\lm} =
P_{\mu\lm} \phi^i_{\mu\lm} =
{1 \over (r-1)!} {\cal L}^{-2}J_{\mu\vkp}J_{\lm\vkp} \phi^i_{\mu\lm} 
= 0 \ . \ \ \ ({\rm on \ shell})
\label{onshell}
\ee
Here, the Companion equation (\ref{jsq}) appears as the covariant 
derivative $\nabla_{\mu}$ acting on $\phi^i_{\mu}$, which explicitly shows 
the general covariance of the Companion equation.
\par

\section{DBI Theory vs. Companion Theory}%
The standard formulation of branes is given by a mapping 
$X^\mu(\tau^i)$ from the $n$-dimensional worldvolume to the
$d$-dimensional target space.
Let us consider the Lagrangian defined with derivatives of $X^\mu(\tau^i)$,
\be
{\cal L}_p = (\det|g_{ij}|)^p \ , \qquad \ \ 
g_{ij} = {\pd X^\mu \over \pd \tau^i} {\pd X^\mu \over \pd \tau^j} \ .
\label{Lp}
\ee
The equation of motion for ${\cal L}_p$ is written, as in \cite{oldpaper}, 
in terms of $g_{ij}$ and the Christoffel symbol
$\Gamma^k_{ij}\,=\, g^{km}\pd_m X^\lm \, \pd_i \pd_j  X^\lm$, 
\be
g^{ij} \pd_i \pd_j X^\mu \,-\, \pd_i X^\mu g^{jk}\Gamma_{jk}^i
\,+\,(2p-1)\pd_i X^\mu g^{ij}\Gamma_{jk}^k \,=\, 0 \ .
\label{greqn}
\ee
Contraction of this equation with $\pd_l X^\mu$ yields
\be
(2p-1)\Gamma_{lk}^k \,=\, (1 - {1 \over 2p})\, {\cal L}_p^{-1} 
\pd_l {\cal L}_p \,=\, 0 \ ,
\label{another}
\ee
and the other terms cancel. 
As in the previous discussion either $p\,=\, {1 \over 2}$, or else 
$\pd_l {\cal L}_p \,=\,0$.
This leaves as equations of motion in all cases
\be
g^{ij} \pd_i \pd_j X^\mu - \pd_i X^\mu g^{jk}\Gamma_{jk}^i = 
( \dl_{\mu\lm} - g^{DBI}_{\mu\lm}) \, g^{ij} \pd_i \pd_j X^\lm
= 0 \ ,
\label{allcase}
\ee
where the degenerate metric 
$g^{DBI}_{\mu\lm} = g^{kl}\pd_k X_\mu \pd_l X_\lm$.
The second equation explicitly shows, via the identity 
$g^{DBI}_{\lm\mu}\pd_l X^\mu = \pd_l X^\lm$, that the number of independent 
equations of motion is $d-n$.
In the following, we will concentrate on the DBI Lagrangian 
${\cal L}_{DBI} = M^n {\cal L}_{1/2}$, with the mass parameter $M$.
\par

\vspace{4mm}

As is known in the particle case ($n=1$), we introduce a field 
$\phi(x)$ as a Hamilton-Jacobi (HJ) function for
${\cal L}_{DBI}$, which gives the canonical conjugate momentum 
of $X^\mu(\tau)$ via the formula $p_\mu(\tau) = \pd_\mu\phi(x=X(\tau))$.
The HJ equation is then obtained from the constraint for $p_\mu$
as $p^2 - M^2 = (\pd_\mu\phi)^2 - M^2 = 0$. 
The generalisation of the HJ formulation to string and brane cases 
has been discussed in \cite{rinke, nambu, knkr, hos1,kastrup,hos2}.
In their paper \cite{hos2}, Hosotani and Nakayama gave a simple
derivation of the HJ equation for general $n > 1$ cases.
They start with the DBI (Nambu-Goto) action for the string ($n=2$);
\be
I_2\,=\, M^2 \int d\tau  d\sig \sqrt{\det|g_{ij}|} 
= M^2\int d\tau d\sig\sqrt{{1 \over 2}
\left(\frac{\pd(X^\mu,X^\nu)}{\pd(\sig,\tau)}\right)^2 } \ .
\label{NG2}
\ee
The covariant momentum tensor $p_{\mu\nu}$ given by
\be
p_{\mu\nu} = {M^2 \over 2} \frac{\frac{\pd(X^\mu,X^\nu)}{\pd(\sig,\tau)}}
{\sqrt{{1 \over 2}\left(\frac{\pd(X^\mu,X^\nu)}{\pd(\sig,\tau)}\right)^2}}
\ , \quad \ \ 
p_{\mu\nu}p^{\mu\nu}\,=\,{M^4 \over 2} \ ,
\label{mom}
\ee
satisfies the equation of motion
\be 
\frac{\pd(p_{\mu\nu},X^\nu)}{\pd(\sig,\tau)}\,=\,0 \ ,
\label{motion}
\ee
which is an alternative form of the equation (\ref{allcase}).
One may think of the solutions for the $X^\mu$ as being functions of 
$\sig,\ \tau$ with $d-2$ additional parameters $\varphi_3,\dots,\varphi_{d}$.
Then $p_{\mu\nu}(\sig,\tau;\varphi_a)$ can be considered as a function 
of the $X^\mu$. 
Following Hosotani and Nakayama, we choose the parameters $\sig,\,
\tau$ in such a way that the area element of the world sheet with fixed 
$\varphi_a$ is
\be
4 M^{-2}p^{\mu\nu}d\sig d\tau\,=\,dX^\mu dX^\nu \ .
\label{area}
\ee
Choosing $\phi^1 =M\sig$ and  $\phi^2=M\tau$, we obtain the
relation between $\pd_i X^\mu$ and $\pd_\mu\phi^i$,
$p_{\mu\nu} = {\tilde J}_{\mu\nu} = \pd_\mu \phi^1 \pd_\nu \phi^2 \,-\,
\pd_\nu \phi^1 \pd_\mu \phi^2$, which gives the HJ equation for strings,
\be
{1 \over 2} {\tilde J}_{\mu\nu} {\tilde J}_{\mu\nu} \,=\,
(\pd_\mu \phi^1)^2(\pd_\nu \phi^2)^2\,-\,(\pd_\mu \phi^1\pd_\mu \phi^2)^2
\,=\, {M^4 \over 4} \ .
\label{hj}
\ee
It is easily seen that the equation of motion (\ref{motion}),
thanks to the Bianchi identity for Jacobians 
$\pd_{[\lm}{\tilde J}_{\mu\nu]} = 0$, is derived from the HJ 
equation as 
\be
{\pd(p_{\mu\nu},X^\nu) \over \pd(\sig,\tau)}
= \pd_\lm {\tilde J}_{\mu\nu}\,{\pd(X^\lm,X^\nu) \over \pd(\sig,\tau)}
= 4 M^{-2} \pd_\lm {\tilde J}_{\mu\nu} \, {\tilde J}_{\lm\nu}
=  M^{-2} \pd_\mu ({\tilde J}_{\lm\nu}{\tilde J}_{\lm\nu}) = 0 \ .
\ee
These results can be generalised straightforwardly to membrane and 
general $p=n-1$-brane cases,
\be
I_n\,=\, M^n \int d^n\tau \sqrt{\det|g_{ij}|} 
= M^n \int d^n\tau \sqrt{{1 \over n!}
\left(\frac{\pd(X^{\mu_1}, \ldots ,X^{\mu_n})}
{\pd(\tau_1,\ldots ,\tau_n)}\right)^2 } \ .
\label{NG}
\ee
The covariant momentum tensor $p_{\vmu} = p_{\mu_1 \cdots \mu_n}$
is set to be equal to the Jacobian for $n$ fields in (\ref{j}),
\be
p_{\vmu} \,=\, {M^{2n} \over n!} {\cal L}_{DBI}^{-1}
\frac{\pd(X^{\mu_1}, \ldots ,X^{\mu_n})}{\pd(\tau_1,\ldots ,\tau_n)}
 \,=\, \eps_{i_1\ldots i_n}\phi^{i_1}_{\mu_1}\ldots\phi^{i_n}_{\mu_n}
\,=\, {\tilde J}_{\vmu} \ ,
\label{pn}
\ee
which leads to the HJ equation, 
\be
{1 \over n!}\,{\tilde J}_{\vmu} \, {\tilde J}_{\vmu} \,=\,
\left({M^n \over n!}\right)^2 \ .
\label{HJ}
\ee
The equation of motion 
$\pd (p_{\mu_1 \mu_2 \cdots \mu_n}, X^{\mu_2}, \ldots, X^{\mu_n})/
\pd (\tau_1, \tau_2, \ldots, \tau_n) = 0$
is solved by the HJ equation and the Bianchi identity 
$\pd_{[\lm} {\tilde J}_{\vmu]} = 0$ as in the string case.
It is interesting to note that, under the relation (\ref{pn}), 
the degenerate DBI metric in (\ref{allcase}) turns out to the 
Companion metric (\ref{deg}),
\be
g^{DBI}_{\mu\lm} = g^{kl}\pd_k X_\mu \pd_l X_\lm = 
{1 \over (n-1)!} \left({n! \over M^n}\right)^2 p_{\mu\vnup}p_{\lm\vnup}
= g_{\mu\lm}(X(\tau)) \ .
\label{gdc}
\ee
\par

\vspace{4mm}

As for the relation between the DBI and the Companion theories,
it is obvious that the HJ equation (\ref{HJ}) in the former is the 
constancy condition of the Lagrangian (\ref{com2}), 
${\cal L} = M^n/n!$, in the latter.
Thus, any field configuration making the Companion Lagrangian
constant solves the DBI equation of motion.
Here, let us consider solutions of the Companion equation 
(\ref{jsq}) in the configuration space of non-zero constant 
Lagrangian, which represent special points of the space to maintain 
the value of the Lagrangian (up to a total derivative) under any variation 
of fields $\phi^i$.
As shown in (\ref{onshell}), the Companion equation is proportional 
to $\nabla_\lm\phi^i_\lm = \pd_\lm g_{\lm\mu}\phi^i_\mu$, which 
would be regarded as a part of the decomposition (\ref{vector}) for
$\pd_\lm g_{\lm\mu}$ into the $n$-dimensional subspace 
${\cal V}_n$,
\be
\nabla_\lm g_{\lm\mu} \,=\, \pd_\lm g_{\lm\mu} \,=\, 
V_i \, \phi^i_\mu 
+ (\dl_{\mu\nu} - g_{\mu\nu}) \pd_{\lm}g_{\lm\nu} \ ,
\label{decom}
\ee
where $V_i = \pd_{\lm}g_{\lm\nu} Y_{i\nu} = 
G_{ij} \nabla_{\nu}\phi^j_{\nu}$.
The second term in the RHS of (\ref{decom}) can be rewritten by using 
another decomposition of 
$\pd_\lm g_{\lm\nu} = \pd_\lm (Y_{j\lm}\phi^j_\nu) 
= \pd_\lm Y_{j\lm}\, \phi^j_\nu + {\cal L}^{-1} \pd_\nu {\cal L}$,
where we used $Y_{j\lm} = {\cal L}^{-1}(\pd {\cal L}/\pd 
\phi^j_{\lm}) $, as
\be
(\dl_{\mu\nu} - g_{\mu\nu}) \, \pd_{\lm}g_{\lm\nu} = 
(\dl_{\mu\nu} - g_{\mu\nu}) \, {\cal L}^{-1} \pd_\nu {\cal L} 
= 0 \ , \quad {\rm if} \ {\cal L} = M^n/n! \ ,
\ee
showing that the subspace of HJ solutions given by the Companion equation
with constant Lagrangian is characterized geometrically by the 
divergence-free condition, 
$\nabla_\lm g_{\lm\mu} = \pd_\lm g_{\lm\mu} = 0$.
\par

\section{Solutions of Companion and HJ equations}%
As has been remarked already, the Companion equation (\ref{jsq}) 
for a single field ($n=1$) in $d$ dimensions takes the form,
\be
\sum_\mu\sum_{\nu\neq\mu}\left((\phi_\nu)^2\phi_{\mu\mu}\,-\, 
\phi_\mu\phi_\nu\phi_{\mu\nu}\right)\label{batsum}\,=\,0 \ ,
\ee
i.e a sum of ${d}\choose{2}$ Bateman equations.
A large class of solutions may be obtained by choosing $d$ 
arbitrary functions $F^\mu(\phi)$ subject to the constraint,
\be
\sum_\mu F_\mu (\phi) x_\mu \,=\, c\,=\, {\rm const.} \ ,
\label{s1}
\ee
and solving  this as an implicit equation for $\phi$. This works 
because this equation implies that
\be
\phi_\mu = \frac{-F_\mu}{ {F_\sig}'x_\sig};\  \  
\phi_{\mu\nu}=\frac{{F_\mu}' F_\nu+{F_\nu}' F_\mu}
{\left({F_\sig}' x_\sig\right)^2}- F_\mu F_\nu
\frac{{F_\lm}'' x_\lm}{\left({F_\sig}'x_\sig \right)^3} \ ,
\label{result}
\ee
where the prime denotes derivatives of $F_\mu$ with respect to 
$\phi$.
These results guarantee that (\ref{batsum}) is satisfied.  
Solutions of this class may be extended to the case of more than 
one field in the following way.
With an ansatz of the form, 
\be
\sum_\mu F^\mu (\phi^1) x_\mu \,=\, c^1 = {\rm const.}\,; \quad
\sum_\mu G_\mu (\phi^2) x_\mu \,=\, c^2 = {\rm const.} \ ,
\label{s2}
\ee
which are solved for $\phi^1$ and $\phi^2$, a similar reasoning 
shows that the Companion equation of motion in an arbitrary 
number of dimensions, which is the sum of ${d}\choose{3}$ 
Universal Field Equations for two fields in three dimensions, admits
this implicit solution, which, in virtue of the covariance property,
can be generalised by replacing the fields $\phi^1$ and $\phi^2$ by 
two arbitrary functions of them.
This class of solutions may be generalised to an arbitrary number 
of fields in an obvious manner.
\par

It can be easily seen that the set of Companion solutions (\ref{s1})
for $n=1$ contains configurations making the Companion 
Lagrangian constant, that is, solutions of the HJ equation.
Assuming the constant $c$ to be non-zero (and rescaled to one)
and $F_\mu(\phi) = \beta_\mu F(\phi)$ with a constant vector 
$\beta_\mu$, we have $F(\phi) = 1/\beta_\mu x_\mu$. 
On the other hand, the form of the function $F(\phi)$ is fixed by the 
condition ${\cal L} = M$, since it breaks the covariance of the Companion
equation under $\phi \rightarrow \phi'$,
\be
{\cal L} \,=\, \sqrt{\phi_\mu\phi_\mu} \,=\,
\sqrt{F_\mu F_\mu \over (F'_\nu x_\nu)^2}
\,=\, \sqrt{\beta^2} {F^2 \over F'} = M \ ,
\ee
where the HJ equation in the base space became a first order differential
equation in the $\phi$-space, which is solved for $F(\phi)$ as 
$-M/\sqrt{\beta^2}\phi$.
This leads to the standard solution of the HJ equation, 
$\phi = p_\mu x_\mu$, where $p_\mu = -M \beta_\mu/\sqrt{\beta^2}$ 
with $p^2 = M^2$.
\par

\vspace{7mm}

In the above particle case, such a standard HJ solution can be 
easily obtained from the HJ equation itself. 
However, in the general $p$-brane cases, the Companion equation
equipped with the remarkable general covariance under field 
redefinitions would become a good starting point to find interesting 
HJ solutions.
It is also intriguing to study the Companion equations for general $(d,n)$
in their own right as a possible extension of the solvable Bateman equation, 
which is now under progress.

\vspace{2.5cm}

\begin{flushleft}
{\bf ACKNOWLEDGEMENT}%
\end{flushleft}
The work of T.U. is partially supported by the Daiwa Anglo-Japanese
Foundation and the British Council.

\newpage

\end{document}